\documentstyle[12pt,world_sci]{article}

\begin{document}
\begin{flushright}
UM-TH-94-26\\
July 1994  \end{flushright}

\title{{\bf MATCHING TO ALL ORDERS AND POWER CORRECTIONS\\
IN HEAVY QUARK EFFECTIVE THEORY}}
\author{M. BENEKE\thanks{Invited talk presented at the 8th Meeting
of the Division of Particles and Fields of the APS, Albuquerque,
New Mexico, August 2 - 6, 1994. To appear in the proceedings.}\\
{\em Randall Laboratory, University of Michigan, Ann Arbor,\\
Michigan 48109, U.S.A.\\}}

\maketitle
\setlength{\baselineskip}{2.6ex}

\begin{center}
\parbox{13.0cm}
{\begin{center} ABSTRACT \end{center}
{\small \hspace*{0.3cm} This talk reports on various aspects of
the divergence of perturbative expansions  in the context of matching
QCD onto heavy quark effective theory. Implications for exclusive
and inclusive decays of heavy mesons are discussed.}}
\end{center}

\section{Introduction}

Heavy quark effective theory (HQET) has become an established tool
in the phenomenology of heavy flavours, with applications expanding
from their original premises of exclusive heavy flavour transitions
into inclusive decays and heavy quark fragmentation\cite{1}. On the
theory side, HQET provides a major example for the framework of
effective field theories (EFT). Thus, given a Green function with
heavy external momenta $p_i=m_Q v+k_i$ and light momenta $q_j$, the
``matching procedure'' extracts the heavy mass dependence in the form
\begin{equation} \label{eq1}
G_{QCD}(p_i,q_j;m_Q;\alpha) = \sum_l\frac{1}{m_Q^l}
\,C_{(l)}\!\left(\frac{m_Q}{\mu};\alpha\right) G^{(l)}_{
HQET}\left(k_i,q_j;\mu;\alpha\right) ,
\end{equation}
when the $p_i$ are close to their mass-shell. The validity of
Eq.~(\ref{eq1}) is established inductively in the number of loops --
i.e. powers of the coupling $\alpha$ -- and to all orders in $l$, where
$\alpha$ and the small off-shellness $|k_i|/m_Q$ are independent
parameters. In practice, one is not so much interested in the matching
of Green functions as in matrix elements between heavy hadrons. In this
case the scale of the off-shellness is provided by the theory itself,
$|k_i|\sim \Lambda_{QCD}$. Since $\Lambda_{QCD}/m_Q\sim
\exp(1/(2\beta_0\alpha(m_Q)))$, the series in power corrections and the
number of loops in the analogue of (\ref{eq1}) are organized in terms
of a single parameter $\alpha(m_Q)$. Leaving the safe grounds of
perturbation theory, one should discuss the presence of power corrections
simultaneously with large order, $\alpha^n$, matching corrections to
the coefficient functions $C_{(l)}$. In fact, the series of these
corrections diverges as $n\rightarrow\infty$ and one source of divergence
originates from low momentum regions, which one would like to factor
into nonperturbative parameters that appear in power corrections.
The summation of the divergent series introduces ambiguities of
the same order of magnitude as these nonperturbative parameters,
which must therefore also be ambiguous. This divergence pattern --
known as renormalons -- and its consequences have a long history\cite{2}
in the context of the short-distance expansion and QCD sum rules. In
this talk I discuss the renormalon phenomenon in HQET and its (ir)relevance
for phenomenology\cite{3,4}.

\section{Renormalon Structure of HQET}

Investigations of large orders in perturbation theory naturally
have to resort to some kind of approximation. Since renormalons are
associated with the integration over logarithms provided by vacuum
polarizations, some insight can be obtained from the restriction
to the class of diagrams generated from insertion of a chain of
fermion bubbles into the low order diagrams. Taking Borel transforms
and defining $u=-\beta_0 t$, factorial divergence of perturbative series
in $\alpha$ translates into singularities of their Borel transforms
in $u$. Singularities at positive $u$ imply non-Borel summability
and an ambiguity in the definition of the sum of the original
divergent series. In the following, $\overline{MS}$ renormalization
in QCD and HQET will be assumed, though $m_Q$ need -- and should --
not coincide with the renormalized mass $m$ of the heavy quark.

The general structure of the Borel transformed version of Eq.~(\ref{eq1})
can be described as follows: The Green functions $G^{(l)}_{HQET}$
in HQET (with operator insertions) are power-like divergent. Explicit
power divergences are absent in dimensional regularization, but they
do not disappear without a trace in $\overline{MS}$. Subtractions
are such that they leave divergent series expansions with non-summable
ultraviolet (UV) renormalons at positive half-integer $u$. It is
natural to associate this divergence with integration over large internal
momenta, $k\gg \mu$, though not straightforward, because integrals
are defined by analytic continuation. The coefficient functions,
$C_{(l)}$, have singularities at positive half-integers, too, which
stem from small internal momenta, $k\ll \mu$. Infrared (IR) renormalons
in coefficient functions cancel with the UV renormalons -- up to the
singularities present already on the l.h.s. of Eq.~(\ref{eq1}). This
cancellation takes place over different orders in the expansion in
$1/m_Q$. Thus, if this expansion is truncated at a certain order,
summation of the perturbative series produces an ambiguous result,
which is removed only by including higher orders in $1/m_Q$.

As an illustration, consider the matching of the inverse propagator
of a heavy quark within the above approximation. The first two
terms of the heavy quark expansion are given by
\begin{equation} \label{eq2}
S^{-1}(p,m;u) = m_Q\left(\frac{m}{\mu};u\right)
- m_{pole}\left(\frac{m}{\mu};u\right)+ C\left(\frac{m_Q}{\mu};u
\right) \star \left(vk\delta(u)-\Sigma_{eff}(vk;u)\right)
+\ldots
\end{equation}
The explicit expressions for the ingredients of Eq.~(\ref{eq2}) lead
to the conclusions\cite{3}:

(a) The pole mass of the heavy quark in the first term on the
r.h.s. has an IR renormalon\cite{5}
at $u=1/2$ when expressed in terms of
$m$. Thus, the pole mass can not be defined to an accuracy better
then $\Lambda_{QCD}$. While this might be expected, the interesting
point is that perturbation theory itself indicates its failure through
its divergence. As a consequence, when the heavy quark expansion
is applied to hadronic parameters, the quantity $\Lambda_{H_Q}
\equiv m_{H_Q}-m_{pole}$, defined as the difference between the
heavy hadron mass and the pole mass of the quark in the heavy quark
limit, contains an ambiguity of order $\Lambda_{QCD}$. Note,
however, that this ambiguity, though of the same order of magnitude
as $\Lambda_{H_Q}$ itself, is not related to bound state effects
contained in $\Lambda_{H_Q}$, but can be traced to the long range
part of the Coulomb field of the quark. Thus, the effect is universal
and obviously cancels in mass differences.

(b) Off mass-shell, the l.h.s. of Eq.~(\ref{eq2}) is non-singular at
$u=1/2$. As anticipated from the general discussion, for $k\not=0$ the
IR renormalon at $u=1/2$ in the pole mass cancels exactly against
an UV renormalon at this position in the self-energy of the static
quark, $\Sigma_{eff}$, computed from the leading term in the HQET
Lagrangian, $\bar{h}_v v\cdot D h_v$. This UV renormalon arises,
since, in contrast with full QCD,
the self-energy of the static quark is linearly UV divergent. This
is nothing but the linear divergence of a static point charge known
from classical electrodynamics, which reappears in HQET, where the
quark mass is considered larger than the UV cutoff.

(c) To reproduce the r.h.s. of Eq.~(\ref{eq2}) from HQET without a
residual mass term, the first term must vanish and the
expansion parameter $m_Q$ has to coincide
with the pole mass. This destroys artifically the cancellation
of renormalon poles, part of which become hidden in the expansion
parameter, which then is not defined beyond perturbation theory.
{}From this point of view it is conceptually advantageous to use
the freedom to add a small residual mass term $-\delta m \bar{h}_v
h_v$ to the effective Lagrangian, such that both $\Sigma_{eff}$
computed from HQET with residual mass and the expansion parameter
$m_Q=m_{pole}-\delta m$ are formally free from an ambiguity due
to a renormalon at $u=1/2$. This can be accomplished either by
$\delta m\propto \mu \sum c_n\alpha(\mu)^n$ $(\mu\ll m_Q)$ with
$c_n$ adjusted to the UV renormalon divergence or a formally ambiguous
$\delta m\propto \mu\exp(1/(2\beta_0\alpha(\mu)))$, adjusted to
compensate the ambiguities of the Borel sums.

\vspace*{0.1cm}
\section{Implications}

Exclusive heavy flavour decays are governed by matrix elements
of the weak current between heavy hadron states. HQET is particularly
effective in restricting the number of independent form factors
in the infinite mass limit and parameterizing the corrections to
this limit. These corrections involve new nonperturbative form
factors and typically the ratio $\Lambda_{H_Q}/m_Q$, which controls
the size of these corrections. Since physical quantities must be
unambiguous, the ambiguity in the definition of $\Lambda_{H_Q}$
implies that the matching corrections in leading order diverge (with
an IR renormalon at $u=1/2$), such that the ambiguity of their sum
compensates the ambiguity in $\Lambda_{H_Q}$, which has been inferred
from the pole mass. It is easy to see that an IR renormalon at
$u=1/2$ will indeed occur. The leading order matching corrections
are conveniently calculated by comparing the current insertions
between on-shell quark states in the full and the effective theory.
In the IR, the integrals behave like $\mbox{d}^4 k/k^4$, but the
coefficient is the same in the full and the effective theory and the
logarithmic IR divergence cancels as it must be. The next term in
the expansion for small $k$, $\mbox{d}^4 k/k^3$, is different, however.
Although this region gives a small and finite contribution to the
coefficient function in first order, it is greatly amplified by
large powers of logarithms, $\ln^n k^2/\mu^2$, in higher orders,
which produces the required divergence of the series.

Thus, the structure of the heavy quark expansion is conceptually
quite similar to the short distance expansion\cite{2}. Power
corrections must be added with care, since the summation of
perturbative corrections, which is never performed in practice,
can produce effects of the same order. In the particular case
of $\Lambda_{H_Q}$ the situation might be more favorable
phenomenologically. This parameter contains the effect of the
light spectators in the heavy hadron, which appears first at this
order and which is clearly not related to renormalon ambiguities.
Given the large value $\Lambda_P\approx 500\,$MeV, favored
for pseudoscalar mesons, one may argue that it is dominated by
the spectator and renormalon effects may be neglected in comparison.

EFT calculations are most conveniently done in $\overline{MS}$.
Since loop integrations run unrestricted over all momenta,
renormalons inevitably appear in the matching corrections.
Alternatively, one might imagine cutting the low momentum
region explicitly from the Feynman integrals, absorbing them
into nonperturbative parameters in higher orders of the
$1/m_Q$-expansion. Disregarding the practical difficulties of
this procedure,
there is a definite drawback: The nonperturbative parameters are
no longer universal and therefore useless (beyond a certain
accuracy). However, strictly within the framework
of EFT, where nonperturbative effects are not calculated but
parameterized, the renormalon phenomenon never constitutes a
difficulty. Indeed, if one accepts the assumption that IR renormalons
in the coefficient functions are in one-to-one correspondence
with (ambiguities of) nonperturbative parameters, one may
eliminate these parameters up to a certain order in $1/m_Q$ in favor of
physical quantites to obtain predictions for other physical
quantities entirely in terms of physical quantities (up to
a certain order in $1/m_Q$). Then, the relation between measurable
quantities will always be free from renormalons up to renormalons
corresponding to a still higher order in $1/m_Q$. (Depending
on the definition of the coupling, it might be necessary to eliminate
the coupling as well.)

The significance of renormalons appears in two respects: First,
when one attempts to calculate the subleading nonperturbative
parameters such as $\Lambda_{H_Q}$, e.g. from QCD sum rules or
lattice gauge theory. In the latter case, the difficulty is
rather profound and appears as explicit power divergences in
the lattice spacing that require ``nonperturbative
subtractions''\cite{6}. Second, the structure of renormalons
serves as a check that IR effects are indeed
correctly parameterized by matrix elements of higher
dimensional operators. As an example, consider the semileptonic
decay width for a $B$ meson. To leading order in $1/m_b$, the
width is naturally proprotional to $G_F^2 m_{b,pole}^5$.
Operator product expansion and HQET predict corrections to
the free quark decay starting\cite{7} from $1/m_b^2$ in apparent
conflict with an ambiguity of order $\Lambda_{QCD}$ from the
IR region in the pole mass. In this case it turns out that the
renormalon in the radiative corrections to the free quark decay
cancels exactly against the one hidden in the pole mass, when
the pole mass is eliminated in favor of a mass parameter that
is not sensitive to the Coulomb tail of the
self-energy\cite{4,5}, implying consistency with the
$1/m_b$-expansion.

\vspace*{0.1cm}
\bibliographystyle{unsrt}

\end{document}